\documentstyle[prb,aps]{revtex}
\begin{document}
\draft
\title{Quasi-two-dimensional hole ordering and dimerized state\\
in the CuO$_2$-chain layers in Sr$_{14}$Cu$_{24}$O$_{41}$}
\author{M. Matsuda}
\address{
The Institute of Physical and Chemical Research (RIKEN),
Wako, Saitama 351-0198, Japan}
\author{T. Yoshihama and K. Kakurai}
\address{
Neutron Scattering Laboratory, ISSP, University of Tokyo,
Tokai, Ibaraki 319-1106, Japan}
\author{G. Shirane}
\address{
Physics Department, Brookhaven National Laboratory,
Upton, New York 11973}
\date{Received 3 August 1998}
\maketitle
\begin{abstract}
Neutron scattering experiments have been performed on
Sr$_{14}$Cu$_{24}$O$_{41}$ which consists of both chains and
ladders of copper ions. We observed that the magnetic excitations
from the CuO$_2$ chain have two branches and that both branches
are weakly dispersive along the $a$ and $c$ axes. The
$\omega$-$Q$ dispersion relation as well as the
intensities can be reasonably described by a random phase
approximation with intradimer coupling between
next-nearest-neighbor copper spins $J$=11 meV, interdimer
coupling along the $c$ axis $J_c$=0.75 meV, and interdimer
coupling along the $a$ axis $J_a$=0.75 meV. The dimer
configuration indicates a quasi-two-dimensional hole ordering,
resulting in an ordering of magnetic Cu$^{2+}$ with
spin-$\frac{1}{2}$ and nonmagnetic Cu, which forms the
Zhang-Rice singlet. We have also studied the effect of Ca
substitution for Sr on the dimer and the hole ordering.
\end{abstract}

\section{Introduction}
Sr$_{14}$Cu$_{24}$O$_{41}$ consists of both two-leg ladders of
copper ions and simple CuO$_2$ chains \cite{mc,siegrist} as shown
in Fig. 1. Numerous experiments showed that the two-leg ladder has
an excitation gap of $\sim$35 meV,
\cite{ecc1,kuma,takigawa,kita,imai}
which is expected theoretically. \cite{dagotto}
An important feature of this compound is
that stoichiometric Sr$_{14}$Cu$_{24}$O$_{41}$ contains hole
carriers. It has been reported that most of the holes are localized in
the chain and some exist in the ladder. \cite{kato,mizuno,osafune}
When Sr$^{2+}$ sites are substituted by Ca$^{2+}$ ions, total
number of the holes in the sample is unchanged but holes in the
chain are transferred from the chain to the ladder
\cite{mizuno,osafune} and the system shows an
insulator-to-metal transition \cite{kato,motoyama}.
Superconductivity was also observed in
Sr$_{0.4}$Ca$_{13.6}$Cu$_{24}$O$_{41}$ below $T_c$=10K
under a high pressure of 3 GPa \cite{uehara}.

In this paper we are only concerned with the magnetic properties in
the chains. Figure 1(a) shows the CuO$_2$ chains in this compound.
The copper ions are coupled by almost 90$^\circ$ Cu-O-Cu bond
along the $c$ axis. Each chain is well isolated from each other along
the $a$ and $b$ axes. As mentioned above, there are localized holes
in the chain. It is expected that the hole spins are localized at oxygen
\cite{mizuno} and couple with copper spins to form Zhang-Rice
singlet \cite{ZR}.
The nonmagnetic Cu sites play an important role to form a dimerized
state in the chain. First, the dimer is formed between
next-nearest-neighbor Cu$^{2+}$ spins along the $c$ axis. The
exchange interaction ($\sim$10 meV) is mediated via a nonmagnetic
ZR singlet. \cite{takigawa,matsu1,ecc2} Then the question is how
the dimers are arranged and interact with each other.
Matsuda $et$ $al.$ \cite{matsu1} interpreted that each dimer is
separated by one ZR singlet along the $c$ axis (Model I) as shown in
Fig. 1(c). They also
interpreted that the excitations at low-$Q$ ($L_{chain}<$0.4)
originate from a dimerized state in the chain and those at high-$Q$
($L_{chain}>$0.6) from a dimerized state due to the interladder
coupling since two branches were observed for the gap excitations at
low-$Q$ but only one branch at high-$Q$.

NMR studies revealed the microscopic properties of the dimerized
state. \cite{takigawa,kita2} It was reported that both magnetic
Cu$^{2+}$ and nonmagnetic ZR singlet exist in the chain. It was
also found that a NMR peak originating from ZR singlet gradually
splits into two peaks below $\sim$200 K. The results were discussed
with two dimer models. One is Model I mentioned above.
This model was supported by Cox $et$ $al.$ by their x-ray
measurements. \cite{cox}
In another model the dimers, which are formed between
next-nearest-neighbor Cu$^{2+}$ spins, are separated by two
nonmagnetic ZR singlets along the $c$ axis (Model II) as shown in
Fig. 1(d). This model is consistent with the results of the electron
diffraction study by Hiroi $et$ $al$. \cite{hiroi} in which the
CuO$_2$ chains show a modulated structure with five times larger
unit cell along the $c$ axis in stoichiometric
Sr$_{14}$Cu$_{24}$O$_{41}$. However, neither models cannot
completely explain the NMR results.

Recently, two unpublished papers \cite{ecc2,regnault} on neutron
scattering experiments came to our attention. Eccleston $et$ $al.$
\cite{ecc2} explained the modulation of the excitations and
intensities along the $c$ axis convincingly by a simple model of an
alternating chain with weak interdimer coupling \cite{barnes}
($J_2/J_1 \sim -0.1$, where $J_1$ and $J_2$ are intradimer and
interdimer coupling, respectively), which is equivalent to Model II.
It is noted that two parallel branches along the $c$ axis are not
separated in the
experiments because of insufficient resolution. Regnault $et$ $al.$
\cite{regnault} briefly reported the measurement of two parallel
branches in a wide range of $Q$ (0.05$\le L_{chain}\le$0.925) along
the $c$ axis, which suggested a weak coupling along the $a$
direction. The specific coupling along the $a$ direction was recently
reported by Cox $et$ $al.$ \cite{cox} in their synchrotron x-ray
study of the charge ordering at low temperatures.

Since dimerized states sometimes occur in spin ($S$)
$\frac{1}{2}$ one-dimensional Heisenberg antiferromagnets
such as the spin-Peierls state, it is quite natural to assume
that the dimerized state in the chain of
Sr$_{14}$Cu$_{24}$O$_{41}$ is also caused by a quantum effect
in $S$=$\frac{1}{2}$ one-dimensional Heisenberg antiferromagnet.
On the other hand, almost isolated magnetic dimers were found
in CaCuGe$_2$O$_6$ \cite{sasago1,zheludev1} even though the
geometrical arrangement of the magnetic moments seems
three-dimensional. It was also claimed that magnetic moments
which are geometrically closest does not necessarily couple
most dominantly in
VODPO$_4 \cdot \frac{1}{2}$D$_2$O \cite{tennant} and
(VO)$_2$P$_2$O$_7$ \cite{garrett} because of the strong
superexchange pathways through a covalently bonded PO$_4$
group. These results suggest that unexpected pathways could
become important in order to completely understand the magnetic
properties of magnetic materials.

We have searched for a simple model of dimers with weak
couplings along both $c$ and $a$ directions. \cite{mm0}
Somewhat surprisingly,
a specific combination of the two couplings, as described by
Leuenberger $et$ $al.$ \cite{leuen} for Cs$_3$Cr$_2$Br$_9$,
produces simple and elegant neutron scattering cross sections
which describe properly the measured dispersion and intensities.
In this paper, we have studied $\omega$-$Q$ dispersion relation
perpendicular to the chain direction in considerable detail.
By applying a random phase approximation (RPA) treatment, we
found that the interdimer coupling along the $a$ axis
($J_a$=0.75 meV) is also important \cite{mm} as well as the
interdimer coupling along the $c$ axis ($J_c$=0.75 meV).
Model cross section will be given after the experimental data are
presented. The dimer arrangement along the
$c$ axis is well described with Model II. The dimer configuration
we found in this study indicates a quasi-two-dimensional hole
ordering, resulting in an ordering of Cu$^{2+}$ and ZR singlet in
the $ac$ plane. Ca substitution for Sr makes the magnetic excitation
peaks broader as in the case of Y substitution \cite{matsu2}
probably because the hole ordering is very sensitive to the hole
number and the long-range dimer formation becomes disturbed.

\section{Experimental Details}
The single crystals of Sr$_{14-x}$Ca$_x$Cu$_{24}$O$_{41}$
($x$=0 and 3) were grown using a traveling solvent floating zone
(TSFZ) method at 3 bars oxygen atmosphere. The dimension of the
cylindrically shaped crystals is about 5 $\times$ 5 $\times$
30 mm$^3$. The effective mosaic of the single crystal is less than
0.4$^\circ$ with the spectrometer condition as described below. The
Sr$_{14}$Cu$_{24}$O$_{41}$ crystal
is the same one as used in Ref. \onlinecite{matsu1}. It is expected
that Sr and Ca are distributed homogeneously in
Sr$_{11}$Ca$_3$Cu$_{24}$O$_{41}$ since
the lattice constants systematically change and the linewidth
of the nuclear Bragg peaks does not change when the ratio of Sr
and Ca is changed. The lattice constants of
Sr$_{14}$Cu$_{24}$O$_{41}$ and
Sr$_{11}$Ca$_3$Cu$_{24}$O$_{41}$ are $a$=11.472 $\AA$
and $c$=27.551 $\AA$ and $a$=11.430 $\AA$ and $c$=27.487
$\AA$ at 15 K, respectively. The lattice constants are consistent
with those obtained with powder samples. \cite{kato}

The neutron scattering experiments were carried out on the
ISSP-PONTA spectrometer installed at the 5G beam hole of the
Japan Research Reactor 3M (JRR-3M) at Japan Atomic Energy
Research Institute (JAERI). The horizontal collimator sequences
were 40$'$-40$'$-S-80$'$-80$'$. The final neutron energy was fixed
at $E_f$=14.7 meV. Pyrolytic graphite (002) was used as
monochromator and analyzer. Contamination from higher-order
beam was effectively eliminated using pyrolytic graphite filters
after the sample. The single crystals were mounted in a closed cycle
refregirator and were oriented in the $(h,0,l)$
scattering plane. As described in Ref. \onlinecite{mc}, there are
three different values for the lattice constant $c$
($c\rm_{universal}$=10 $\times$ $c\rm_{chain}$=7 $\times$
$c\rm_{ladder}$). Since we will show the magnetic and
structural properties in the chain, $c\rm_{chain}$ will be used to
express Miller indices.

\section{Experimental results}
\subsection{Magnetic excitations from
Sr$_{14}$Cu$_{24}$O$_{41}$ chain}
\subsubsection{Low temperature dispersion}
Figure 2 shows the typical neutron inelastic spectra at
($H$,0,$L$) in Sr$_{14}$Cu$_{24}$O$_{41}$ measured at 15 K.
One or two distinct excitation peaks are observed in the $(H,0,L)$
scattering plane as in the $(0,K,L)$ scattering plane which was
previously reported in Ref \cite{matsu1,matsu2}. The solid lines at
$(2,0,-0.20)$, $(2.75,0,-0.20)$, $(2,0,-0.75)$, and $(3,0,-0.75)$ are
fits to two Gaussians and those at $(2,0,-0.70)$ and $(2.75,0,-0.70)$
are the fits to a single Gaussian. The excitation
peak positions are changed when $H$ or $L$ is changed,
meaning a dispersion along the $a$ and $c$ axes. The width of the
excitation peaks varies at different $Q$ positions purely due to the
resolution focussing effect. At $(3,0,-0.75)$, where resolution
focussing is almost perfect, the peak width has a resolution limited
value of $\sim$1 meV full width at half maximum. The observed
excitation energies and energy-integrated intensities at (2,0,$L$) and
(3,0,$L$) (0.1$\le L\le$0.85) are plotted in Figs. 3 and 4, respectively.
The closed circles represent the data from the magnetic peaks with
lower excitation energy and the open circles from the ones with
higher excitation energy. The solid and broken lines are the results of
model calculations, which will be presented in the next section.
Both at $(2,0,L)$ and $(3,0,L)$ the dispersion relations are similar
and has a periodicity of 0.2 r.l.u. along the $c$ axis which is
consistent with the results by Eccleston $et$ $al$. \cite{ecc2}
and Regnault $et$ $al$. \cite{regnault}
Actually the dispersion relation is almost identical to
the one observed in the $(0,K,L)$ zone. \cite{matsu1}
The most interesting feature is that the intensities and the ratio of
the two peaks are changed with $H$ as well as $L$.
The intensities of the excitation peaks with lower excitation energy
are more intense than those with higher excitation energy at
$(2,0,L)$ ($L<$0.25). On the other hand, the intensities of the
excitation peaks with higher excitation energy are more
intense than those with lower excitation energy at $(2,0,L)$
($L>$0.6). The intensities at $(2,0,L)$ are almost identical to
those observed in the $(0,K,L)$ zone. \cite{matsu1}
The intensities at $(3,0,L)$ show completely opposite behavior from
those at $(2,0,L)$. It is noted that the intensities in
Ref. \onlinecite{ecc2}
is the sum of intensities from the two excitations.

Figures 5 and 6 show the observed excitation energies and
energy-integrated intensities at $(H,0,-0.2)$ and $(H,0,-0.7)$
(2$\le H\le$4), respectively.
The closed circles represent the data from the magnetic peaks with
lower excitation energy and the open circles from the ones with
higher excitation energy. The solid and broken lines are the results
of model calculations, which will be presented in the next section.
There also exists a dispersion along the $a$ axis with a periodicity of
2 r.l.u. The two dispersion curves along the $a$ axis do not run
parallel as ones along the $c$ axis but cross at
$H=\frac{5}{2}$ and $\frac{7}{2}$. The intensities of the
excitation peaks with lower excitation energy are larger than those
with higher excitation energy in a range of 2$<H<$2.5 and
3.5$<H<$4 at $(H,0,-0.2)$. On the other hand, in a range of
2.5$<H<$3.5 the intensities of the excitation peaks with higher
excitation energy are larger than those with lower excitation energy.
The intensities at $(H,0,-0.7)$ show completely opposite behavior
from those at $(H,0,-0.2)$.
Thus, both the dispersion and the intensities depend on $H$ and $L$,
suggesting that there are nonnegligible magnetic correlations along
the $a$ axis in addition to those along the $c$ axis.

\subsubsection{Model Hamiltonian}
In order to analyze the observed dispersion relation and
energy-integrated intensities we used a model Hamiltonian for the
dimers formed between next-nearest-neighbor Cu$^{2+}$ spins
which are weakly coupled along the $a$ and $c$ axes as shown in
Fig. 7. The Hamiltonian involves three Heisenberg Hamiltonians
with intradimer coupling $J$, interdimer coupling in the same chain
$J_c$, and interdimer coupling between the adjacent chains $J_a$:

\begin{eqnarray}
\hat{H} = J \sum_{<i,j>}
\mbox{\boldmath $S$}_i\cdot \mbox{\boldmath $S$}_j +
J_c \sum_{<i,j>'}
\mbox{\boldmath $S$}_i\cdot \mbox{\boldmath $S$}_j +
J_a \sum_{<i,j>''}
\mbox{\boldmath $S$}_i\cdot \mbox{\boldmath $S$}_j 
\label{Ham}
\end{eqnarray}
Here, $<i,j>$ and $<i,j>'$ are nearest-neighbor and
next-nearest-neighbor spins in the same chain, respectively.
$<i,j>''$ is nearest-neighbor spins between the adjacent chains. This
type of dimers, namely weakly coupled dimers, can be well
described with RPA. In fact, the magnetic excitations in
Cs$_3$Cr$_2$Br$_9$ \cite{leuen} and BaCuSi$_2$O$_6$
\cite{sasago2} are successfully described with RPA treatment. The
dispersion relation is given by \cite{leuen}

\begin{eqnarray}
\omega^{\rm acoustic/optic}(\mbox{\boldmath $q$}) &=&
[J^2+J\cdot R(T)\left( J_c\Gamma _c(\mbox{\boldmath $q$})
\pm J_a|\Gamma _a(\mbox{\boldmath $q$})|\right) ]^{1/2},
\nonumber \\
R(T)&=&n_0-n_1=\frac{1-{\rm exp}(-J/T)}{1+3{\rm exp}(-J/T)},
\nonumber \\
\Gamma_c(\mbox{\boldmath $q$})&=&2{\rm cos}(2\pi l),
\nonumber \\
\Gamma_a(\mbox{\boldmath $q$})&=&2{\rm cos}(\pi h)\cdot
{\rm exp}(i2\pi l\delta),
\label{omega}
\end{eqnarray}
where $n_0$, $n_1$, and $\delta$ are thermal populations of the
singlet ground state and of the first excited triplet and the distance
along the $a$ axis between nearest-neighbor spins at the adjacent
chains as shown in Fig. 7. The solid and broken lines in
Figs. 3(a)-6(a)
represent the calculated dispersion relation of the acoustic and
optic modes, respectively. Equation (2) with $J$=11 meV,
$J_a$=0.75 meV, and $J_c$=0.75 meV reproduces the observed
data remarkably well.
In this dispersion relation, the averaged excitation energy, the band
width of each excitation mode, and the energy difference between
the acoustic and optic modes are related with $J$, $J_c$, and
$J_a$, respectively. $J$ and $J_c$ obtained in this experiment
are consistent with those obtained by Eccleston $et$ $al$. \cite{ecc2}
The sign of $J_a$ is consistent with that in
La$_{14-x}$Ca$_x$Cu$_{24}$O$_{41}$ which shows
a long range magnetic ordering. \cite{matsu3,matsu4}

RPA treatment gives the dynamic structure factor \cite{leuen} as
the following.

\begin{eqnarray}
S(\mbox{\boldmath $Q$},\omega) &\propto &
F^2(\mbox{\boldmath $Q$})\cdot R(T)\cdot J\cdot
[1-{\rm cos}(\mbox{\boldmath $Q$}\cdot \mbox{\boldmath $R$})]
\nonumber \\
&\times & \left [ [1+
{\rm cos}(\mbox{\boldmath $\rho$}\cdot
\mbox{\boldmath $\tau$}+\phi)]
\frac{1}{\omega^{\rm acoustic}(\mbox{\boldmath $q$})}
\delta \left( \omega -\omega^{\rm acoustic}(\mbox{\boldmath $q$})
\right) \right. \nonumber \\
&+& \left.
[1-{\rm cos}(\mbox{\boldmath $\rho$}\cdot
\mbox{\boldmath $\tau$}+\phi)]
\frac{1}{\omega^{\rm optic}(\mbox{\boldmath $q$})}
\delta \left( \omega -\omega^{\rm optic}(\mbox{\boldmath $q$})
\right) \right]
\label{Sqw}
\end{eqnarray}
where $\phi = 2\pi l\delta$,
$\mbox{\boldmath $Q$}$=$\mbox{\boldmath $q$}$ +
$\mbox{\boldmath $\tau$}$ ($\mbox{\boldmath $\tau$}$ is a
reciprocal wave vector), and \mbox{\boldmath $R$} is the
vector connecting individual spins within a dimer. The factor 
$[1-{\rm cos}(\mbox{\boldmath $Q$}\cdot \mbox{\boldmath $R$})]$
can be seen in the structure factor for isolated dimers.
The acoustic and optic modes can be distinguished by the factor
$[1\pm{\rm cos}(\mbox{\boldmath $\rho$}\cdot
\mbox{\boldmath $\tau$}+\phi)]$.
The solid and broken lines in Figs. 3(b)-6(b) represent the
energy-integrated intensities of the acoustic and optic modes
calculated using Eq. (3), respectively. $J$, $J_a$, and $J_c$ are
fixed at the values mentioned above. $\delta$ is fixed at 0.77 which
was determined by x-ray measurements at 50 K. \cite{cox} Only
adjustable parameter is the scale factor. The same scale factor was
used for the intensities at (2,0,$L$), (3,0,$L$), $(H,0,-0.2)$, and
$(H,0,-0.7)$. The energy-integrated intensities are described with
the RPA theory reasonably well.

\subsubsection{Temperature dependence}
Figure 8 shows constant-$Q$ scans at $(2,0,-0.1)$ and $(2,0,-0.6)$
as a function of temperature. The solid lines are the fits to two
Gaussians. In the fitting the higher excitation energy at $(2,0,-0.1)$
and the lower excitation energy at $(2,0,-0.6)$ are fixed, which is
confirmed with more intense excitation peaks at $(3,0,-0.1)$ and
$(3,0,-0.6)$. It was assumed that the peak width
and the intensity ratio of the two excitations are temperature
independent. Temperature dependence of excitation energies and
energy-integrated intensities are plotted in Fig. 9. The lower
excitation energy at $(2,0,-0.1)$ becomes higher and the higher
excitation energy at $(2,0,-0.6)$ becomes lower with increasing
temperature. Furthermore, the
intensities decrease with increasing temperature. 
Since there is a thermal factor $R(T)$ in Eqs. (2) and (3), the RPA
calculation predicts the temperature dependence of the dispersion
relation and the energy-integrated intensities.
The dispersion becomes flatter with increasing temperature as shown
in the inset of Fig. 9. because the interdimer couplings become
negligible due to large thermal fluctuations and the dimers behave
like isolated dimers. The intensity is proportional to $R(T)$ which
decreases with increasing temperature.
The solid lines in Fig. 9 are the results of calculations using Eq. (2)
and (3). All the parameters $J$, $J_a$, $J_c$, $\delta$, and scale
factor are fixed at the values as determined above. The temperature
dependences are fitted with the RPA calculation reasonably
well although the observed intensity decreases more quickly than the
calculated intensity does. This indicates that other factors in addition
to thermal fluctuations disturb the dimers and the ordering of
Cu$^{2+}$ and ZR singlet. A gradual destruction of the
ordering with increasing temperature suggested by NMR
\cite{takigawa} and x-ray measurements \cite{cox} might cause the
behavior.

\subsubsection{Discussion}
Now we discuss the relation between magnetic dimerization
and structural distortion. As mentioned in Sec. I, Cox $et$ $al.$
\cite{cox} performed synchrotron x-ray measurements on
Sr$_{14}$Cu$_{24}$O$_{41}$ and observed superlattice Bragg
peaks at (0,0,$l$/4) and (0,0,$l$/2) which originate from a
structural distortion in the chain. This lattice distortion is probably
related with the NMR results \cite{takigawa,kita2} since the
superlattice intensity also decreases gradually with increasing
temperature. It should be pointed out that the x-ray experiments
properly detected the coupling along the $a$ axis but there is some
disagreement along the $c$ axis. As mentioned above, the structural
distortion has correlations with two and four times periodicities
along the $c$ axis although magnetic dimers have correlations with
five times periodicity.
A discrepancy between magnetic and structural correlations was
also reported in the spin-Peierls state in CuGeO$_3$. \cite{hirota}
The magnetic excitations have a minimum energy at
(0,1,$\frac{1}{2}$) but a superlattice reflection at
($\frac{1}{2}$,1,$\frac{1}{2}$).
An important point is that the x-ray results are for the ground state
and neutron results are for the excited state.

We have shown that the dimer model shown in Fig. 7 explains the
magnetic excitations reasonably well. The dimerized state is ascribed
to the ordering of Cu$^{2+}$ and ZR singlet which originates from
an ordering of the localized holes.
Hole ordering or charge ordering occurs in various 3$d$ transition
metal oxides. A quasi-two-dimensional hole ordering of the doped
holes was reported in La$_2$NiO$_{4+\delta}$ \cite{tranquada2}
and La$_{2-x-y}$Nd$_x$Sr$_y$CuO$_4$ \cite{tranquada}.
In the phase a static ordering of antiferromagnetic stripe is separated
by charge ordered domain walls. In the latter compound the ordered
phase can be described as a stripe ordering of Cu$^{2+}$ and ZR
singlet.
Manganese oxides, for example La$_{1-x}$Ca$_x$MnO$_3$, also
show a charge ordering of Mn$^{3+}$ and Mn$^{4+}$.
\cite{wollan} In this system spin, charge, and orbital \cite{good}
degrees of freedom are closely related, which results in interesting
phenomena such as colossal magnetoresistance. \cite{schiffer}
NaV$_2$O$_5$, which consists of two-leg ladders of vanadium
ions, has been studied extensively since it shows a singlet ground
state below 34 K. \cite{isobe} The phase transition
was first considered to be a conventional spin-Peierls transition.
\cite{fujii} But the detailed inelastic neutron scattering experiment
using single crystal sample revealed that the magnetic excitations in
the dimerized phase could not be understood by a simple
one-dimensional dimerized model. \cite{yoshihama} NMR
experiment also indicated a charge ordering of V$^{4+}$ and
V$^{5+}$ at the transition. \cite{ohama} Stimulated by these
experimental findings, new approaches to understand the origin of 
the phase transition have been made theoretically.
\cite{seo,mostovoy} It is claimed that the transition is indeed closely
related with a charge ordering of V$^{4+}$ and V$^{5+}$ ions.
These various properties suggest that an intimate
connection between hole/charge ordering and magnetic ordering is
a common feature in the strongly correlated 3$d$ transition metal
oxides. Theoretical studies to explain the hole ordering and the
dimerized state in Sr$_{14}$Cu$_{24}$O$_{41}$ are highly
desirable.

\subsection{Substitution effect in
Sr$_{14-x}$Ca$_x$Cu$_{24}$O$_{41}$ ($x$=3)}
We have also studied the substitution effect on the dimerized state.
The dimerized state in the chain is most stable in pure
Sr$_{14}$Cu$_{24}$O$_{41}$. When Y is substituted for Sr,
the number of the holes is decreased and the number of the
Cu$^{2+}$ spins is increased. It was reported that Y substitution
makes the magnetic excitation peaks broader. \cite{matsu2}
As mentioned in Sec. I, the Ca
substitution for Sr also reduces the number of the holes in the
chain gradually. \cite{kato,mizuno,osafune} Therefore, the Ca
substitution is expected to affect the dimerized state in the chain.

Figure 10 shows typical inelastic neutron spectra at (3,0,$L$)
measured at 8 K in Sr$_{11}$Ca$_3$Cu$_{24}$O$_{41}$. The
magnetic excitation peaks become broader with Ca substitution,
which is similar to the case of Y substitution. \cite{matsu2} It is
noted that two excitation branches are hardly resolved. However,
the excitation energies are similar to those in
Sr$_{14}$Cu$_{24}$O$_{41}$. The solid lines at $(3,0,-0.15)$,
$(3,0,-0.30)$, and $(3,0,-0.35)$ are fits to two Gaussians and that at
$(3,0,-0.75)$ is a fit to a single Gaussian by assuming that
the dispersion relation is the same as that in
Sr$_{14}$Cu$_{24}$O$_{41}$. The spectra are reasonably
described with the simple model.
As mentioned in Sec. III-A, the averaged excitation
energy, the band width of each excitation mode, and the energy
difference between the acoustic and optic modes are related with
$J$, $J_c$, and $J_a$, respectively.
This indicates that the coupling constants $J$, $J_c$, and $J_a$ are
almost unchanged with Ca substitution although the dimerized state
becomes unstable. This behavior would be explained as the
following. The hole ordering sensitively depends on the hole
number. Accordingly, the long-range dimer formation becomes
disturbed with Ca or Y substitution.

In Sr$_{14-x}$Y$_x$Cu$_{24}$O$_{41}$ system, magnetic
excitations are remarkably affected with Y substitution.
\cite{matsu1,matsu2} In the $x$=0.25 sample the excitation peaks
become broader although the dispersion relation is almost unchanged.
This result suggests that $J$, $J_c$, and $J_a$ are almost unchanged
although the hole ordering is disturbed with a small amount of Y.
In the $x$=1 sample the excitation peaks become much broader,
suggesting that the hole ordering becomes more unstable.
An interesting behavior is that the averaged energy of the two
excitations becomes reduced to $\sim$9 meV,
indicating that $J$ is decreased in
Sr$_{13}$Y$_1$Cu$_{24}$O$_{41}$.
The changes of the bandwidth of each excited state, the difference
in energy between two excited state, and the periodicity of the
dispersion relation are difficult to be determined because the peaks
are too broad to be resolved as two excitations.
A puzzling feature is that $J$ is decreased in
Sr$_{13}$Y$_1$Cu$_{24}$O$_{41}$ although the lattice constant
$c$, which affects the exchange constant, is almost independent of Y
concentration. \cite{kato2}

When the holes are removed further, the hole ordering cannot be
observed anymore and a long-range magnetic
ordering appears. \cite{matsu3,matsu4,carter} The CuO$_2$ chains
in the hole removed Sr$_{14}$Cu$_{24}$O$_{41}$ can be
considered as ferromagnetic spin chains which are weakly coupled
with antiferromagnetic interchain coupling along the $a$ and $b$
axes.

In conclusion,
neutron scattering experiments have been performed on
Sr$_{14}$Cu$_{24}$O$_{41}$ in order to study the dimerized
state in the CuO$_2$ chains. The $\omega$-$Q$ dispersion relation
as well as the integrated intensities can be reasonably described by
RPA calculation with $J$=11 meV, $J_c$=0.75 meV, and
$J_a$=0.75 meV. The dimer configuration indicates a
quasi-two-dimensional hole ordering, which results in an ordering
of magnetic Cu$^{2+}$ with $S=\frac{1}{2}$ and nonmagnetic ZR
singlet. The hole ordering and the dimerized state are most stable in
Sr$_{14}$Cu$_{24}$O$_{41}$. The Ca substitution
for Sr sites makes the excitation peak broader probably because the
hole ordering becomes unstable and the long-range dimer
formation becomes disturbed.

\section*{Acknowledgments}
We would like to thank D. E. Cox, R. S. Eccleston, H. Eisaki,
K. Katsumata, S. M. Shapiro, M. Takigawa, and A. Zheludev for
stimulating discussions. This
study was supported in part by the U.S.-Japan Cooperative
Program on Neutron Scattering operated by the United States
Department of Energy and the Japanese Ministry of Education,
Science, Sports and Culture and by a Grant-in-Aid for Scientific
Research from the Japanese Ministry of Education,
Science, Sports and Culture. Work at Brookhaven National
Laboratory was carried out under Contract No.
DE-AC02-98CH10886, Division of Material Science, U.S.
Department of Energy.

\begin{figure}
\caption{Structure of the CuO$_2$ chains (a) and the Cu$_2$O$_3$
ladders (b) in Sr$_{14}$Cu$_{24}$O$_{41}$. The lower part of the
figure shows two dimer models as described in Section I: (c) Model
I and (d) Model II. The squares denote Zhang-Rice singlet sites.
Figure from ref. 18.}
\label{fig1}
\end{figure}

\begin{figure}
\caption{Typical inelastic neutron spectra at ($H$,0,$L$)
at 15 K in Sr$_{14}$Cu$_{24}$O$_{41}$. The solid lines are
the results of fits to a single Gaussian or two Gaussians.}
\label{fig2}
\end{figure}

\begin{figure}
\caption{Observed and calculated energies (a) and intensities (b) at
(2,0,$L$) measured at 15 K in Sr$_{14}$Cu$_{24}$O$_{41}$.
The solid and broken lines, which are fits to Eqs. (2) and (3),
represent the acoustic and optic modes, respectively.}
\label{fig3}
\end{figure}

\begin{figure}
\caption{Observed and calculated energies (a) and intensities (b) at
(3,0,$L$) measured at 15 K in Sr$_{14}$Cu$_{24}$O$_{41}$.
The solid and broken lines, which are fits to Eqs. (2) and (3),
represent the acoustic and optic modes, respectively.}
\label{fig4}
\end{figure}

\begin{figure}
\caption{Observed and calculated energies (a) and intensities (b) at
$(H,0,-0.2)$ measured at 15 K in Sr$_{14}$Cu$_{24}$O$_{41}$.
The solid and broken lines, which are fits to Eqs. (2) and (3),
represent the acoustic and optic modes, respectively.}
\label{fig5}
\end{figure}

\begin{figure}
\caption{Observed and calculated energies (a) and intensities (b) at
$(H,0,-0.7)$ measured at 15 K in Sr$_{14}$Cu$_{24}$O$_{41}$.
The solid and broken lines, which are fits to Eqs. (2) and (3),
represent the acoustic and optic modes, respectively.}
\label{fig6}
\end{figure}

\begin{figure}
\caption{A proposed model for the dimerized state and the ordering
of Cu$^{2+}$ and ZR singlet in the $ac$ plane.}
\label{fig7}
\end{figure}

\begin{figure}
\caption{Inelastic neutron spectra at $(2,0,-0.1)$ ($T$=15 and 75 K)
and $(2,0,-0.6)$ ($T$=15 and 100 K)
in Sr$_{14}$Cu$_{24}$O$_{41}$. The solid and broken lines are
the results of fits to two Gaussians.}
\label{fig8}
\end{figure}

\begin{figure}
\caption{Temperature dependence of the energy-integrated
intensity and energy of the gap excitations at $(2,0,-0.1)$ and
$(2,0,-0.6)$ in Sr$_{14}$Cu$_{24}$O$_{41}$. The solid lines are
fits to Eqs. (2) and (3). The inset shows the calculated dispersion at
15 and 100 K.}
\label{fig9}
\end{figure}

\begin{figure}
\caption{Typical inelastic neutron spectra at (3,0,$L$) at 8 K in
Sr$_{11}$Ca$_3$Cu$_{24}$O$_{41}$. The solid lines are
the results of fits to a single Gaussian or two Gaussians.}
\label{fig10}
\end{figure}


\begin{references}
\bibitem{mm0}The dispersion along the $b$ axis is almost flat and
the intensities along the $b$ axis also monotonically decrease with
increasing $Q$ due to the magnetic form factor as reported in
Ref. \onlinecite{matsu1}.
\bibitem{mm}In Ref. \onlinecite{matsu1} it was mentioned that the
dispersion for the magnetic excitations at low-$Q$ is almost flat
along the $a$ axis. The mistake was made because the
measurements were performed mostly at $H$=integer.
\end{references}
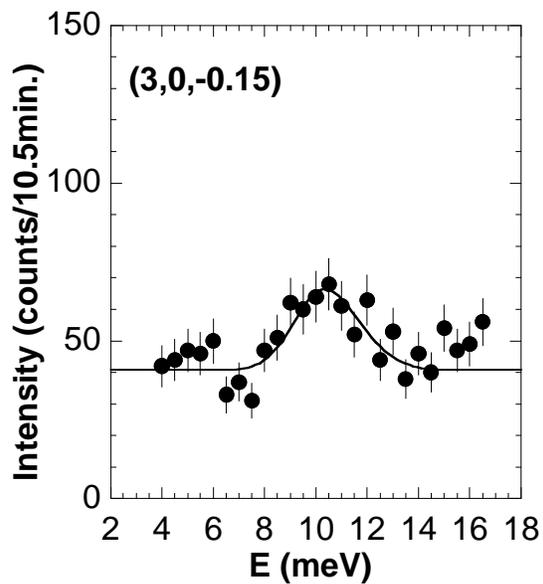
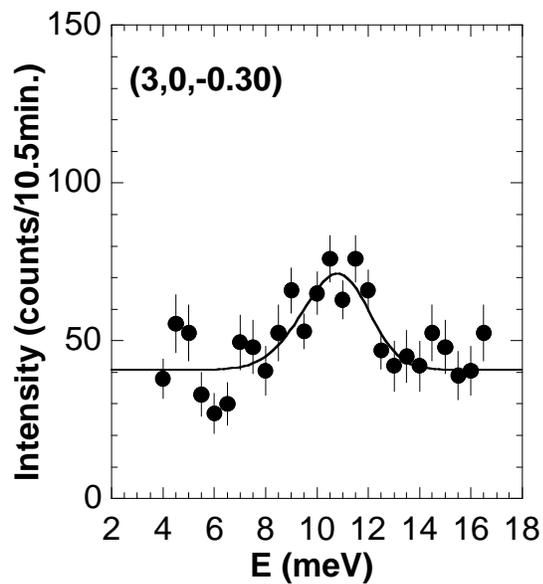
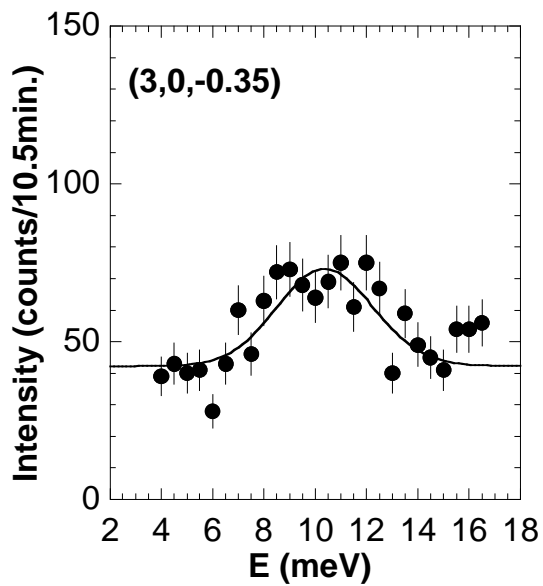
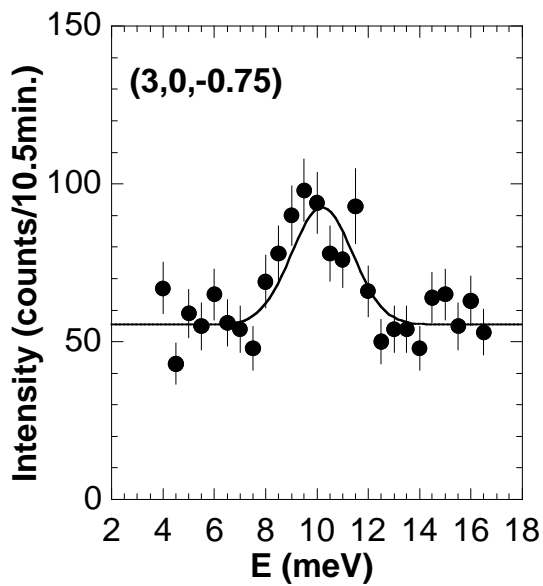
\end{document}